
\magnification=1200
\baselineskip=13pt
\overfullrule=0pt
\tolerance=100000

{\hfill \hbox{\vbox{\settabs 1\columns
\+ UR-1404 \cr
\+ ER-40425-851\cr
\+ hep-th/9501095\cr
}}}

\bigskip
\bigskip
\baselineskip=18pt

\centerline{\bf Properties of an Alternate Lax Description of the KdV
Hierarchy}

\vfill

\centerline{J. C. Brunelli}
\medskip
\centerline{and}
\medskip
\centerline{Ashok Das}
\medskip
\medskip
\centerline{Department of Physics and Astronomy}
\centerline{University of Rochester}
\centerline{Rochester, NY 14627, USA}
\vfill

\centerline{\bf {Abstract}}

\medskip
\medskip

We study systematically the Lax description of the KdV hierarchy in terms of an
operator which is the geometrical recursion operator. We formulate the Lax
equation for the $n$-th flow, construct the Hamiltonians which lead to
commuting
flows. In this formulation, the recursion relation between the conserved
quantities follows naturally. We give a simple and compact definition of all
the Hamiltonian structures of the theory which are related through a power law.
\vfill
\eject

\bigskip
\noindent {\bf 1. {Introduction}}
\medskip

The reduction of the nonlinear KdV (Korteweg-de Vries) equation to a system of
linear equations by Lax [1]
gave a new impetus to the study of integrable models [2-4].
The generalization of the Lax pair to particular pseudo-differential operators
has yielded many interesting results associated with various integrable systems
including the Hamiltonian structures associated with such systems [5]. As an
example, let us note that for the KdV hierarchy we have
$$
\eqalign{
L=&\partial^2+{1\over 6}u\cr
B_n=&a_n\left(L^{2n+1\over2}\right)_+\cr
}\eqno(1)
$$
where $a_n$ is a constant and $(\quad)_\pm$
will refer to terms which contain non-negative (negative)
powers of $\partial$ in a pseudo-differential operator.

On the other hand, we know that the KdV hierarchy is bi-Hamiltonian [6] with
the
two Hamiltonian structures given by
$$
\eqalign{
{\cal D}_1=&\partial\cr
{\cal D}_2=&\partial^3+{1\over3}(\partial u+u\partial)
}\eqno(2)
$$
In such a case, one can define a natural geometrical structure which is the
recursion operator as the Lax operator, namely,
$$
\eqalign{
R={\cal D}_2{\cal D}_1^{-1}=&\partial^2+{1\over3}u+
{1\over3}\partial u\partial^{-1}\cr
=&\partial^2+{2\over3}u+
{1\over3}u_x\partial^{-1}\cr
}\eqno(3)
$$
This operator enjoys special properties, e.g., it relates the conserved
quantities of the system recursively and gives all the Hamiltonian structures
of the theory through a power relation. (This happens only if the Nijenhuis
torsion tensor associated with $R$ vanishes and this is known to be true for
the KdV hierarchy [6-8].)

It would be, therefore, interesting to study the Lax formulation of the KdV
hierarchy through this ``geometrical'' operator. We note that the formulation
of the KdV equation as a Lax equation with the Lax operator $R$
is well known [9-11].
However, a systematic study of the entire KdV hierarchy through this operator
has not been undertaken to the best of our knowledge -- mainly because the
standard formulation is quite successful. In this note, we undertake a
systematic study of the Lax formulation of the KdV hierarchy
in terms of the operator $R$ and bring out all its properties. In
sec. 2, we recapitulate all the relevant facts about the KdV hierarchy. In sec.
3, we present our formulation with all its properties and present our
conclusions in sec. 4.
\bigskip
\noindent {\bf 2. {Properties of the KdV Hierarchy}}
\medskip

In this section, we will recapitulate briefly the relevant properties of the
KdV hierarchy [3,5]. The $n$-th equation in the KdV hierarchy has the form
($u$ denotes the dynamical variable.)
$$
{\partial u\over \partial t_{n+1}}=K_{n+1}(u)
=\left({\cal D}_1G_{n+1}(u)\right)
=\left({\cal D}_2G_{n}(u)\right)\qquad n=0,1,2,\dots\eqno(4)
$$
where ${\cal D}_1$ and ${\cal D}_2$ are the Hamiltonian structures defined
in (2) and
$$
G_n(u)={\delta H_n\over\delta u}\eqno(5)
$$
with $H_n$ representing the $n$-th  conserved charge of the system. The first
few conserved quantities have the explicit form
$$
\eqalign{
H_0=&3\int dx\,u\cr
H_1=&{1\over2}\int dx\,u^2\cr
H_2=&{1\over2}\int dx\,\left({1\over3}u^3-u_x^2\right)
}\eqno(6)
$$
{}From (4) it is clear that the Hamiltonians of the theory are related
recursively through the recursion operator defined in (3) as (The bracket
defines the action of the operators.)
$$
{\delta H_{n+1}\over\delta u}=
\left(R^\dagger{\delta H_n\over\delta u}\right)\eqno(7)
$$
where the formal adjoint of $R$ is defined [12] to be
$$
\eqalign{
R^\dagger&=\partial^2+{1\over3}u+{1\over3}\partial^{-1}u\partial\cr
         &=\partial^2+{2\over3}u-{1\over3}\partial^{-1}u_x\cr
}\eqno(8)
$$

It is also clear from (2) and (4) that for any flow (any value of $n$)
$$
\eqalign{
K_n(u)=&\left(\partial G_n(u)\right)\cr
K_{n+1}=&(R\,K_n(u))\cr
}\eqno(9)
$$
Similarly, it is known that the $n$-th Hamiltonian structure of the hierarchy
is related to the first through the power relation
$$
{\cal D}_n=R^{n-1}{\cal D}_1\qquad n\ge1\eqno(10)
$$
Finally, let us note, without going into details [1,3,7,9], that if we define
the Frechet derivative (these are operators)
$$
(K_{n+1})'\equiv M_{n+1}\eqno(11)
$$
where $(F)'$ denotes the Frechet derivative of $F$ (We use this notation to
avoid confusion with $F'$ which arises later and would denote derivative of $F$
with respect to $x$.), then we can show that
$$
{\partial R\over \partial t_{n+1}}=[M_{n+1},R]\eqno(12)
$$
which also yields the equations of the KdV hierarchy.
With all these basics, we are now ready to study systematically the Lax
formulation of the KdV hierarchy with the recursion operator $R$.

\bigskip
\noindent {\bf 3. {Properties of the Alternate Lax Formulation}}
\medskip

Let us note here a few properties of the recursion operator $R$. We have
$$
\eqalign{
R^\dagger=&\,\partial^2+{1\over3}u+{1\over3}\partial^{-1}u\partial=
\partial^{-1}R\,\partial\cr
R^{1/2}=&\,\partial+{1\over3}u\partial^{-1}-{1\over 18}u^2\partial^{-3}+
{1\over9}uu_x\partial^{-4}-
{1\over9}\left({1\over2}u_x^2+uu_{xx}-{1\over6}u^3\right)\partial^{-5}
+\cdots\cr
}\eqno(13)
$$
Here the subscript denotes a derivative with respect to the corresponding
variable. It follows now that
$$
\eqalign{
\hbox{Tr}\,R^{1/2}=&\int dx\,\hbox{Res}\,R^{1/2}={1\over3}\int dx\,u\cr
\hbox{Tr}\,R^{3/2}=&\int dx\,\hbox{Res}\,R^{3/2}={1\over6}\int dx\,u^2\cr
\hbox{Tr}\,R^{5/2}=&\int dx\,\hbox{Res}\,R^{5/2}=
{5\over18}\int dx\,\left({1\over3}u^3-u_x^2\right)\cr
}\eqno(14)
$$
where ``Res'' stands for the coefficient of the term $\partial^{-1}$ in the
expression. It is clear that these are nothing other than the first three
Hamiltonians (up to normalization constants) given in (6). In fact, it is
straightforward to see that the properly normalized Hamiltonians can be written
as
$$
H_n={9\over 2n+1}\hbox{Tr}\,R^{{2n+1}\over2}\qquad n=0,1,2,\dots\eqno(15)
$$
To see this, let us note from equations (7) and (13) that
$$
\eqalign{
{\delta H_n\over\delta u}=&\left(R^\dagger{\delta H_{n-1}\over\delta u}\right)=
\left((R^\dagger)^n{\delta H_{0}\over\delta u}\right)\cr
=&9(R^\dagger)^n{\delta\ \over\delta u}\hbox{Tr}\left(R^\dagger\right)^{1/2}
={9\over 2n+1}{\delta\ \over\delta u}
\hbox{Tr}\left(R^\dagger\right)^{{2n+1\over2}}\cr
=&{9\over 2n+1}{\delta\ \over\delta u}
\hbox{Tr}\,R^{{2n+1\over2}}\cr
}\eqno(16)
$$
where we have used $(13)$ as well as the cyclicity properties of the trace and
which leads to the identification in (15).

The expression for the Hamiltonians in (15), in terms of
the Lax operator, is exactly
the same as the conventional case [5]. Therefore, it is natural to expect a Lax
equation for the dynamical equations of the conventional form
$$
{\partial R\over\partial t_{n+1}}=\left[\left(
R^{{2n+1\over2}}\right)_+,R\right]\qquad n=0,1,2,\dots\eqno(17)
$$
This would, of course, ensure the commutativity of the different flows.
However, let us note that unlike the conventional Lax operator, $L$, the
recursion operator contains both positive and negative powers of $\partial$ and
as such it is not clear, a priori, whether an equation of the form (17)
is even consistent. From the structure of $R$ in (3), we note that for (17) to
be consistent, we must have
$$
\left(R^{{2n+1}\over2}\right)_+=\partial P\eqno(18)
$$
where $P$ contains only nonnegative powers of $\partial$. But it is not at all
clear that this is indeed the case. We note here that the time evolution of $R$
is given by (12) and from the definitions in (9) and (11) we see
that (these are operators)
$$
M_{n+1}=\partial(G_{n+1})'\eqno(19)
$$
which has the right structure necessary for consistency. If we can
show that
$$
M_{n+1}=\left(R^{{2n+1}\over2}\right)_+\eqno(20)
$$
then, we would have proved the form of the Lax equation to coincide with (17).
In fact, we can check explicitly from the form of $M_n$ and $R$ that
$$
\eqalign{
M_1=&(R^{1/2})_+\cr
M_2=&(R^{3/2})_+\cr
}\eqno(21)
$$
However, (20) has to be shown in general for any $n$ to prove (17).

Let us note that we can write [13]
$$
R^{{2n+1}\over2}=(R^{1/2})^{2n+1}=\sum_{i\le 2n+1}\partial^i a_i(n)=
\sum_{i\le 2n+1}b_i(n)\partial^i\eqno(22)
$$
where the coefficients $a_i(n)$ and $b_i(n)$ are related. In particular
$$
\eqalign{
b_{-1}(n)=&a_{-1}(n)\cr
b_{-2}(n)=&a_{-2}(n)-a'_{-1}(n)\cr
b_{-3}(n)=&a_{-3}(n)+a''_{-1}(n)-2a'_{-2}(n)\cr
}\eqno(23)
$$
and so on. We also note
that
$$
(R^{1/2})^\dagger=-\partial^{-1}R^{1/2}\partial\eqno(24)
$$
which leads to
$$
\left((R^{1/2})^{2n+1}\right)^\dagger\partial^{-1}
=-\partial^{-1}(R^{1/2})^{2n+1}\eqno(25)
$$
Using the representation (22) in (25), we obtain
$$
\sum_{i\le 2n+1}(-1)^i a_i(n)\partial^{i-1}=-\sum_{i\le
2n+1}\partial^{i-1}a_i(n)\eqno(26)
$$
Comparing the coefficients of $\partial^{-1}$ on both sides, we then get
$$
a_0(n)=-a_0(n)=0\eqno(27)
$$
{}From the representation in (22) we now see that
$$
\left(R^{{2n+1}\over2}\right)_+=\sum_{i=1}^{2n+1}\partial^ia_i(n)=
\partial P\eqno(28)
$$
as required in (18).

Next, we note that
$$
\left(R^{{2n+3}\over2}\right)^\dagger=
R^\dagger\left(R^{{2n+1}\over2}\right)^\dagger=
\left(R^{{2n+1}\over2}\right)^\dagger R^\dagger\eqno(29)
$$
which upon using the representation in (22) leads to two relations
$$
\eqalign{
\sum_{i\le 2n+3}(-1)^i a_i(n+1)\partial^i=&
\left(\partial^2+{2\over3}u-{1\over3}\partial^{-1}u_x\right)
\sum_{i\le 2n+1}(-1)^i a_i(n)\partial^i\cr
\sum_{i\le 2n+3}(-1)^i a_i(n+1)\partial^i=&
\sum_{i\le 2n+1}(-1)^i a_i(n)\partial^i
\left(\partial^2+{2\over3}u-{1\over3}\partial^{-1}u_x\right)\cr
}\eqno(30)
$$
These equations, along with (27) give (after some algebra)
$$
\eqalign{
a_{-2}(n)-a'_{-1}(n)&=0\cr
a_{-3}(n)-2a'_{-2}(n)+a''_{-1}(n)&=
-{1\over3}\left(\partial^{-1}u_xa_{-1}(n)\right)\cr
}\eqno(31a)
$$
or
$$
\eqalign{
b_{-2}(n)&=0\cr
b_{-3}(n)&=-{1\over3}\left(\partial^{-1}u_xb_{-1}(n)\right)\cr
}\eqno(31b)
$$
Thus, this analysis shows that for any  nonnegative integer, $n$,
$$
\eqalign{
R^{{2n+1}\over2}=&\left(R^{{2n+1}\over2}\right)_+
+\left(R^{{2n+1}\over2}\right)_-\cr
=&\partial P(n)+b_{-1}(n)\partial^{-1}
-{1\over3}\left(\partial^{-1}u_xb_{-1}(n)\right)\partial^{-3}+\cdots\cr
}\eqno(32)
$$
This has the consequence that for any $n$, we have
$$
\eqalign{
\hbox{Res}\,R^{{2n+1}\over2}\partial=&0\cr
\hbox{Res}\,\partial^{-1}\left(R^{{2n+1}\over2}\right)_+=&\hbox{Res}\,
\partial^{-1}R^{{2n+1}\over2}=0\cr
\hbox{Res}\,R^{{2n+1}\over2}\partial^2=&
-{1\over3}\left(\partial^{-1}u_x\hbox{Res}\,R^{{2n+1}\over2}\right)\cr
}
\eqno(33)
$$

Given these relations, we can now easily show from the definition of the
Hamiltonian in (15) that
$$
\eqalign{
G_n=&{\delta H_n\over\delta u}=3\left(\hbox{Res}\,R^{{2n-1}\over2}\right)\cr
\hbox{Res}\,R^{{2n-1}\over2}=&\hbox{Res}\,RR^{{2n-3}\over2}=
\left(\left(\partial^2+{2\over3}u-{1\over3}\partial^{-1}u_x\right)
\hbox{Res}\,R^{{2n-3}\over2}\right)\cr
}
\eqno(34)
$$
so that
$$
{\delta H_n\over\delta u}=\left(R^\dagger{\delta H_{n-1}\over\delta u}\right)
\eqno(35)
$$
Namely, the recursion relation of equation (7) is built into our definition of
the Hamiltonian in (15). We note here that the conventional Lax operator, $L$,
does not lead to the recursion relation in any simple way.

Next, let us note that using the properties in (33), we can show that
$$
\left(R^{{2n+1}\over2}\right)_+=R\left(R^{{2n-1}\over2}\right)_+
+2\left(\partial\hbox{Res}\,R^{{2n-1}\over2}\right)
+\left(\hbox{Res}\,R^{{2n-1}\over2}\right)\partial\eqno(36)
$$
On the other hand, from the relations in (11) and (34), it can be shown in a
straightforward manner that
$$
M_{n+1}=R\,M_n
+2\left(\partial\hbox{Res}\,R^{{2n-1}\over2}\right)
+\left(\hbox{Res}\,R^{{2n-1}\over2}\right)\partial\eqno(37)
$$
Comparing the recursion relations (36) and (37) and noting the identification
in (21), we conclude that for any nonnegative integer value of $n$
$$
M_{n+1}=\left(R^{{2n+1}\over2}\right)_+\eqno(38)
$$
This proves that the $n$-th equation in the KdV hierarchy is given by the Lax
equation
$$
{\partial R\over\partial t_{n+1}}=\left[\left(
R^{{2n+1\over2}}\right)_+,R\right]\qquad n=0,1,2,\dots\eqno(39)
$$
as suspected. The commutativity of the flows now follows in the standard
fashion.

Next, we turn to the question of the Hamiltonian structures in this alternate
Lax formulation of the KdV hierarchy. We note from the structure of $R$ in (3)
that we can define a dual to be of the form
$$
V=\partial^{-1}v_{-1}+\partial^{-2}v_{-2}\eqno(40)
$$
However, since the coefficients of $\partial^0$ and $\partial^{-1}$ in $R$ are
related, we expect $v_{-1}$ and $v_{-2}$ to be functionally related for
consistency. This can be easily seen from the Lax equation
$$
{\partial R\over\partial t_{n+1}}=\left[\left(
R^{{2n+1\over2}}\right)_+,R\right]=\left[R,\left(
R^{{2n+1\over2}}\right)_-\right]\eqno(41)
$$
where we note from (22) and (31) that
$$
\left(R^{{2n+1\over2}}\right)_-=\partial^{-1}a_{-1}(n)+\partial^{-2}a_{-2}(n)
+\cdots\eqno(42)
$$
with
$$
\eqalign{
a_{-1}(n)=&\hbox{Res}\,R^{{2n+1}\over2}={1\over3}
{\delta H_{n+1}\over\delta u}\cr
a_{-2}(n)=&a'_{-1}(n)\cr
}\eqno(43)
$$
Thus, consistency would require us to choose a dual of the form
$$
V=\partial^{-1}v+\partial^{-2}v_x\eqno(44)
$$
A standard analysis (see ref. [14] for details)
then shows that with the Hamiltonian structure defined as
$$
\{F_Q(R),F_V(R)\}_1={2\over9}\hbox{Tr}\,R[V,Q]\eqno(45)
$$
we can write (41) in the Hamiltonian form
$$
{\partial F_Q(R)\over\partial t_{n+1}}=\{F_Q(R),H_{n+1}\}_1\eqno(46)
$$
where we have defined the linear functionals as
$$
F_V(R)=\hbox{Tr}\, RV={2\over3}\int dx\,u(x)v(x)\eqno(47)
$$

To obtain the higher order Hamiltonian structures, we note that if we use (35)
in the standard analysis of the Lax equation, we can show that with the
definition of the Hamiltonian structure
$$
\{F_Q(R),F_V(R)\}_n={2\over9}\hbox{Tr}\,R[((R^\dagger)^{n-1}V),Q]
\qquad n\ge1\eqno(48)
$$
the Lax equation (41) can be written in the Hamiltonian form
$$
{\partial F_Q(R)\over\partial t_{n+1}}=\{F_Q(R),H_{n+1}\}_{1}=
\{F_Q(R),H_{n}\}_{2}=\cdots=\{F_Q(R),H_{1}\}_{n+1}\quad n=0,1,2,\dots
\eqno(49)
$$
Here we have defined the dual operator $((R^\dagger)^{n-1}V)$ as
$$
((R^\dagger)^{n-1}V)\equiv\partial^{-1}((R^\dagger)^{n-1}v)+
\partial^{-2}((R^\dagger)^{n-1}v_x)\eqno(50)
$$

The $n$-th structure constructed from (49) is easily seen to satisfy (using
properties such as the first relation in (13))
$$
{\cal D}_n=R^{n-1}{\cal D}_1 \qquad n\ge1\eqno(51)
$$
as expected. We note that unlike the conventional Lax description, here we have
a simple and compact definition of any Hamiltonian structure of the theory
which leads directly to the power law relation (10) among them.

\bigskip
\noindent {\bf 4. {Conclusion}}
\medskip

We have systematically studied the Lax formulation of the KdV hierarchy where
we have used the natural, geometrical recursion operator as the Lax operator.
By using special properties associated with this operator, we have shown that
the $n$-th flow of the KdV hierarchy continues to be described in the standard
Lax form in terms of the new Lax operator $R$. The Hamiltonians of the system
are given by traces of odd half integer powers of this operator and this leads
to the commutativity of various flows. However, in this  formulation, the
recursion relation between the conserved quantities follows in a simple manner.
This is not the case in the description of the system in terms of the
conventional Lax operator. We have
also shown that in this alternate Lax description, all the Hamiltonian
structures of the theory can be defined in a simple and compact way and lead to
the power law relation among them. Once again, this is an improvement over the
conventional Lax representation. Generalization of our study to other
integrable systems remains an open question.

\bigskip
\noindent {\bf Acknowledgements}
\medskip

This work was supported in part by the U.S. Department of Energy Grant No.
DE-FG-02-91ER40685. J.C.B. would like to thank CNPq, Brazil, for
financial support.

\vfill\eject

\noindent {\bf {References}}
\bigskip

\item{1.} P. D. Lax, Comm. Pure Appl. Math. {\bf 21}, 467 (1968);
ibid {\bf 28}, 141 (1975).

\item{2.} L.D. Faddeev and L.A. Takhtajan, ``Hamiltonian Methods in
the Theory of Solitons'' (Springer, Berlin, 1987).

\item{3.} A. Das, ``Integrable Models'' (World Scientific, Singapore,
1989).

\item{4.} M.J. Ablowitz and P.A. Clarkson, ``Solitons, Nonlinear
Evolution Equations and Inverse Scattering'' (Cambridge, New York, 1991).

\item{5.} L. A. Dickey, ``Soliton Equations and Hamiltonian Systems'' (World
Scientific, Singapore, 1991).

\item{6.} F. Magri, J. Math. Phys. {\bf 19}, 1156 (1978). The KdV equation is
really a tri-Hamiltonian system. See, for example, J. C. Brunelli and A. Das,
``KdV and NLS Equations as Tri-Hamiltonian Systems'', University of Rochester
preprint UR-1391 (1994) (also hep-th/9410165).

\item{7.} S. Okubo and A. Das, Phys. Lett. {\bf 209B}, 311 (1988);
A. Das and S. Okubo, Ann. Phys. {\bf 190}, 215 (1989).

\item{8.} A. Das and W.-J. Huang, J. Math. Phys. {\bf 31}, 2603 (1990).

\item{9.} H. H. Chen, Y. C. Lee and C. S. Liu, Phys. Scr. {\bf 20}, 490 (1979).

\item{10.} F. Magri, in ``Nonlinear Evolution Equations and Dynamical
Systems'',
M. Boiti, F. Pempinelli and G. Soliani, eds., Lecture Notes in Physics, No. 120
(Springer, New York, 1980).

\item{11.} P. J. Olver , ``Applications of Lie Groups to Differential
Equations'', Graduate Texts in Mathematics, Vol. 107 (Springer, New York,
1986).

\item{12.} E. Date, M. Kashiwara, M. Jimbo and T. Miwa, in ``Nonlinear
Integrable Systems - Classical Theory and Quantum Theory'', ed. M. Jimbo and T.
Miwa (World Scientific, Singapore, 1983).

\item{13.} B.A. Kupershmidt, Commun. Math. Phys. {\bf 99}, 51 (1985).

\item{14.} A. Das and W.-J. Huang, J. Math. Phys. {\bf 33}, 2487 (1992).

\end